\documentclass[prd]{revtex4} 

\usepackage{amsmath} 
\usepackage{amssymb} 
\usepackage{graphicx} 
 
\newcommand{\re}[1]{(\ref{#1})} 
\newcommand{\ve}[1]{\mathbf{#1}}

\begin{document} 
 
\title{ Vacuum creation of massive vector bosons and 
its  application to a conformal cosmological model} 
 
\author{D.B. Blaschke$^{1,2}$, 
A.V. Prozorkevich$^3$, and S.A. Smolyansky$^3$} 
 
\affiliation{$^1$Fakult\"at f\"ur Physik, Universit\"at 
Bielefeld, D-33615 Bielefeld, Germany \\
$^2$ Bogoliubov Laboratory for Theoretical Physics,
Joint Institute for Nuclear Research, 141980 Dubna, Russia \\ 
$^3$Physical Department of Saratov State University,
410026 Saratov, Russia}

\begin{abstract} 
In the simple model of a massive vector field in flat 
space-time, we derive a kinetic equation of non-Markovian 
type, describing the vacuum pair creation  under the action of 
external fields of different nature. We use for this 
aim the non-perturbative methods of kinetic theory in 
combination with a new element  when the transition of the 
instantaneous quasiparticle representation is realized 
within the oscillator (holomorphic) representation. We 
study in detail the process of vacuum creation of vector 
bosons generated by a time-dependent boson mass in 
accordance with a conformal-invariant scalar-tensor 
gravitational theory and its cosmological application. It 
is indicated that the choice of the equation of state (EoS) 
of the Universe allows to obtain a number density of 
the vector bosons that is sufficient to explain the 
observed number density of photons in the cosmic microwave 
background radiation. It is shown that the vector boson gas 
created from the vacuum is in a strong non-equilibrium state and 
corresponds to a cold dust-like EoS. 
\end{abstract} 
 
\maketitle 
 
\section{Introduction} 

The vacuum creation of massive vector bosons in intense 
fields of different nature is widely discussed in the 
literature \cite{Nik}-\cite{Most} and it is particularly 
interesting since the massive vector field is the simplest 
example of a quantum field theory with higher spin \cite{Kruglov} 
and can be used to model aspects of gauge field theories.
To mention examples: the massive vector bosons of the standard 
model play an important role in different physical problems  
in cosmology (see, e.g., \cite{DB,Grib82}), the massive vector 
field can be considered as a simplified version of QCD 
\cite{Skalozub}, and the $\rho$ meson as a ``massive photon''
plays a key role in the diagnostics of hot and dense 
nuclear matter in heavy-ion collisions via its decay into 
lepton pairs \cite{rapp}.

Different methods have been used for the investigation 
of the vacuum quantum effects for vector particles, 
e.g., the imaginary time method \cite{Popov}, the 
Bogoliubov canonical transformation method \cite{Nik,Most}, and 
the oscillator representation \cite{DB}. The most attention 
has been devoted to the massive vector particles with gyromagnetic 
ratio  $g=2$. The probability of pair creation of a 
vector field with arbitrary $g$ in the constant 
electromagnetic field has been considered in \cite{Popov}. 
The generalization to arbitrary time dependent electric 
field has been performed in \cite{Most} using the method of 
diagonalization of the Hamiltonian, where problems have been 
encountered  in treating pair production by electric fields for the
case $g=2$. This was unexpected in view of a successful 
evaluation of the Lagrange function for a constant field in the
one-loop approximation \cite{VT}. 
 
In the present work we give a kinetic description of the vacuum 
creation of charged massive vector bosons under the 
influence of a time dependent spatially uniform electric 
field of arbitrary polarization. We also consider the 
possibility of a time dependent mass which represents a new 
independent mechanism of vacuum particle production. 
 
The use of kinetic methods allows one to obtain a rather 
general solution of the non-perturbative problem for an 
arbitrary time dependence of the strong external fields. 
The non-perturbative approach is particularly appropriate 
for fastly changing fields such as, for example, in the case 
of time dependent vector boson masses in the vicinity of 
the cosmological singularity, see Sect. \ref{UNI}. 
 
The construction of a kinetic theory of vacuum particle 
creation on a dynamical basis requires the  time dependent 
quasi-particle representation (QPR) \cite{Smol,OR}. We use 
the oscillator representation for this aim \cite{OR} as the 
most effective instrument for the derivation of dynamical 
equations in the QPR in Sect.\ref{QPR}.  We introduce here 
two types of QPR: the complete one (based on the full 
diagonalization of all physical quantities in the Fock and 
spin spaces) and the incomplete one which lets the spin 
projection uncertain. Further in Sect. \ref{KE} we use these 
results for the derivation of the kinetic equations (KE). A 
new feature of the obtained system of KE is the presence of 
a tensor distribution function in a rotating coordinate 
system with the orientation defined by a time dependent 
kinematic momentum that results in a new type of 
non-Markovian processes. A significant simplification is 
achieved when the non-Markovian effects are neglected. The 
case of the absence of an electric field is considered in 
detail when the vacuum creation is caused entirely by the 
time dependence of the mass. The system of KE splits  into 
separate equations for the transverse and the longitudinal 
components which can be investigated numerically. As an 
application we reinvestigate in Sect. \ref{UNI} the creation 
of massive vector bosons in the early Universe within a 
conformal invariant scalar-tensor theory of gravitation as 
suggested earlier by Pervushin and collaborators 
\cite{DB,Perv02}. In this approach the time dependence of 
the scalar field entails a cosmological evolution of all 
particle masses which, according to Hoyle and Narlikar may 
serve as an explanation for the cosmological red shift 
alternative to the Hubble expansion. In the present 
approach, we are able to remove the singularity in the 
density of the produced longitudinal vector bosons which 
has been reported previously \cite{DB}. We present a 
solution of the KE for a toy model where the time 
dependence of the scalar field is given and show that the 
density of vector bosons created  in the early Universe 
corresponds to the number density of cosmic microwave 
background (CMB) photons. As it will be shown in 
Sect. \ref{UNI}, the massive vector boson-anti-boson gas 
created from the vacuum is a cold one. It leads to a coarse grained 
pressure that proves to be approximately zero, see Sect. \re{EOS}. 
At the same time, the energy density grows 
at large time scales proportional to the particle mass. 
That leads to a dust-like equation of state (EOS) of 
massive vector boson - anti-boson gas, in distinction from 
the static stiff EOS of the usual gas of massive vector 
bosons \cite{Zeld61}. Finally in Sect. \ref{SUM} we 
summarize and present the conclusion. 
 
We use the metric $g^{\mu\nu}=diag(1,-1,-1,-1)$ and natural 
units $\hbar =c =1$. 

\section{The quasi-particle representation}\label{QPR} 
 
We consider the vacuum creation of charged massive vector 
bosons in the flat Minkowski space-time by the action of two 
mechanisms: (i) a time variation of the boson mass $m(t)$ and 
(ii) the action of some classical spatially homogeneous time-dependent 
   electric field with 4-potential (in the Hamilton gauge) 
\begin{equation}\label{eq1} 
    A^{\mu}(t) = \left (0,A^{1}(t),A^{2}(t),A^{3}(t) \right ), 
\end{equation} 
where the corresponding field strength is $\vec E=-\dot{\vec A}$, 
and the overdot denotes the time derivative. 
 
Thus, the field can be considered either as an external field, or as 
a result of the mean field approximation, based on the 
substitution of the quantized electric field $\tilde A^k(t)$ with 
its mean value $\langle\,\tilde A^k(t)\,\rangle =A^k(t)$, where 
the symbol $\langle\,\ldots\,\rangle $ denotes some averaging 
operation. The time dependence of the vector boson mass can be 
interpreted as a result of the coupling to an average Higgs field. 
In the kinetic theory, the consideration of fluctuations leads to 
the collision integrals \cite{Smol03}. Thus, the mean field 
approximation corresponds to the neglect of dissipative effects. 
 
We will restrict ourselves to the simplest version of the theory 
with the Lagrangian
\begin{equation}\label{eq2} 
    \mathcal{L}(x) = - D_{\mu}^{*} u_\nu^{*} D^{\mu} 
     u^\nu + m^{2}u_\nu^* u^\nu~, 
\end{equation} 
where $ D_{\mu} = \partial_{\mu} + i e A_{\mu} $, and $e$ is the 
charge of the vector field including its sign. 
Eq. (\ref{eq2}) leads to the equation of motion 
\begin{equation}\label{eq2a} 
    (D_{\mu}D^{\mu} + m^2) u_\nu = 0 
\end{equation} 
 with the additional constraint 
\begin{equation}\label{add} 
  D_\mu u^\mu=0. 
\end{equation} 

The transition to the QPR can be realized in different 
ways, e.g. by means of the time-dependent Bogoliubov 
transformation \cite{Grib94}, or with the help of the 
holomorphic (oscillator) representation (OR) \cite{OR}. We 
choose the OR being the simpler method. The OR can be 
introduced in the spatially homogeneous case and it is 
based on the replacement of the canonical momentum by the 
kinematic one $\ve{p}\to\ve{P}=\ve{p}-e\ve{A}$ in the 
dispersion law of the free particle 
$\omega(\ve{p},t)=\sqrt{m^2(t)+\ve{P}^2}$ to be used in the standard 
decomposition of the free field operators  and momenta in 
the discrete momentum space \cite{BS} 
\begin{align} 
 u_\mu (x) & = \ \frac{1}{\sqrt{V}}\sum\limits_{\ve{p}} 
 \frac{1}{\sqrt{2 \omega(\ve{p},t)}}\, \ {\rm e}^{i \ve{p} \, \ve{x}} 
\ \left \{ a^{(-)}_\mu(\ve{p},t) + b^{(+)}_\mu(-\ve{p},t) 
 \right \}, \nonumber \\ 
 \pi_\mu (x) &= -\frac{i}{\sqrt{V}} \sum\limits_{\ve{p}} \, 
 \sqrt{2\omega(\ve{p},t)}\, {\rm e}^{-i \ve{p} \, \ve{x}} 
\left \{ a^{(+)}_\mu(\ve{p},t) - b^{(-)}_\mu (-\ve{p},t) 
 \right \}\label{or}, 
\end{align} 
where $V=L^3$ and  $p_i= (2\pi/L)n_i  $ with an integer $n_i$ for 
each $i=1,2,3$. The substitution of the field operators (\ref{or}) 
into the Hamiltonian 
\begin{equation}\label{hamx} 
  H= -\int d\ve{x} \left(\pi^*_\mu \pi^\mu+ 
  \ve{D}^* u_\mu^*\ve{D}u^\mu +m^2 u_\mu^*u^\mu \right) 
\end{equation} 
brings it at once to a diagonal form in the Fock space which 
corresponds to the QPR 
\begin{equation}\label{hamp} 
  H= -\sum\limits_{\ve{p}} \,\omega(\ve{p},t) \biggl[a_\mu^{(+)}(\ve{p},t) 
  a^{(-)\mu}(\ve{p},t)+ b_\mu^{(-)}(-\ve{p},t) 
  b^{(+)\mu}(-\ve{p},t)\biggr]. 
\end{equation} 
However, this quadratic form is not positively defined. In order 
to exclude the $\mu=0$ component with the help of the additional 
condition \re{add}, it is necessary to derive the equations for 
the amplitudes $a^{\pm},b^{\pm} $. 
 
Substituting the Eqs. \re{or}  into the Hamiltonian equations 
\begin{equation}\label{hameq} 
  \dot{u}_\mu=\frac{\delta H}{\delta \pi^\mu} = -\pi^*_\mu,\qquad 
  \dot{\pi}_\mu=-\frac{\delta H}{\delta  u^\mu} = m^2u^*_\mu - 
  \ve{D}^*\ve{D}^*u^*_\mu, 
\end{equation} 
we find the Heisenberg-type equation of motion for the 
time-dependent creation and annihilation amplitudes 
\begin{eqnarray}\label{he1} 
\dot{a}^{(\pm)}_\mu (\ve{p},t)=\frac12\Delta 
(\ve{p},t)b_\mu^{(\mp)}(-\ve{p},t) \pm i\omega(\ve{p},t) 
a^{(\pm)}_\mu(\ve{p},t),\nonumber\\ \dot{b}^{(\pm)}_\mu 
(-\ve{p},t)=\frac12\Delta (\ve{p},t)a_\mu^{(\mp)}(\ve{p},t) \pm 
i\omega(\ve{p},t) b^{(\pm)}_\mu(-\ve{p},t), 
\end{eqnarray} 
where 
\begin{equation}\label{delta} 
  \Delta(\ve{p},t)=\frac{\dot{\omega}(\ve{p},t)}{\omega(\ve{p},t)}. 
\end{equation} 
Analogous equations were obtained in the work \cite{OR} for 
the case of scalar QED on the basis of the principle of least 
action. 
 
Thus, the Hamiltonian formalism in the OR leads to the exact 
equations of motion \re{he1} for the creation and annihilation 
operators of quasi-particles, depending on the "natural" 
representation of the quasi-particle energy $\omega(\ve{p},t)$ in 
the external field \re{eq1}. 
 
The additional conditions \re{add} may be transformed now with the 
help of Eqs. \re{he1} to the following form ($i=1,2,3$) 
\begin{equation}\label{add2} 
\omega(\ve{p},t) a_0^{(\pm)}(\ve{p},t)=P_i 
a_i^{(\pm)}(\ve{p},t),\qquad \omega(\ve{p},t) 
b_0^{(\pm)}(-\ve{p},t)=-P_i b_i^{(\pm)}(-\ve{p},t). 
\end{equation} 
These equations allow to exclude the $\mu=0$ component in the 
Hamiltonian \re{hamp} with the result 
\begin{multline}\label{18} 
H=\sum\limits_{\ve{p}} 
\,\omega(\ve{p},t)\left\{a_i^{(+)}(\ve{p},t)a^{(-)}_i(\ve{p},t)+ 
b_i^{(-)}(-\ve{p},t)b^{(+)}_i(-\ve{p},t)\right. -\\ \left. 
-\frac{1}{\omega^2(\ve{p},t)} \left[ \left(P_i 
a_i^{(+)}(\ve{p},t)\right) \left(P_k a_k^{(-)}(\ve{p},t)\right)+ 
\left(P_i b_i^{(-)}(-\ve{p},t)\right) \left(P_k 
b_k^{(+)}(-\ve{p},t)\right)\right]\right\}. 
\end{multline} 
 
The next step is the additional diagonalization of the quadratic form 
\re{18} by means of the linear transformations \cite{BS} 
\begin{eqnarray}\label{lin} 
\ve{a}^{(\pm)}(\ve{p},t)&=& E\boldsymbol{\alpha}^{(\pm)}(\ve{p},t) 
\equiv\ve{e}_1 \alpha^{(\pm)}_1(\ve{p},t) +\ve{e}_2 
\alpha^{(\pm)}_2(\ve{p},t)+ \ve{e}_3 \frac{\omega}{m} 
\alpha^{(\pm)}_3(\ve{p},t), 
\nonumber\\ 
\ve{b}^{(\pm)}(-\ve{p},t) &=& 
E\boldsymbol{\beta}^{(\pm)}(-\ve{p},t) \equiv\ve{e}_1 
\beta^{(\pm)}_1(-\ve{p},t) 
 +\ve{e}_2  \beta^{(\pm)}_2(-\ve{p},t)+ 
\ve{e}_3 \frac{\omega}{m} \beta^{(\pm)}_3(-\ve{p},t), 
\end{eqnarray} 
where $[\ve{e}_1(\ve{p},t),\ve{e}_2(\ve{p},t),\ve{e}_3(\ve{p},t)]$ 
determine the local rotating basis built on the vector $\ve{e}_3 = 
\ve{P}/|P|$. These real unit vectors form the triad, 
\begin{equation}\label{triad} 
    e_{ik}e_{jk} = e_{ki} e_{kj} =\delta_{ij}, \qquad e_{ik} = 
    (\ve{e}_i)_k \,. 
\end{equation} 
The presence of the factor $\omega/m$ in the non-unitary matrix $E$ in 
Eq. \re{lin}, leads to a violation of the unitary equivalence between 
the $(a,b)$ and $(\alpha,\beta)$ representations. 
 
The transformation \re{lin} leads to the positively defined 
Hamiltonian 
\begin{equation}\label{hd} 
  H= \sum\limits_{\ve{p}}\,\omega(\ve{p},t)\biggl[\alpha_i^{(+)}(\ve{p},t) 
   \alpha_i^{(-)}(\ve{p},t) 
  +\beta_i^{(-)}(-\ve{p},t)\beta_i^{(+)}(-\ve{p},t)\biggr]. 
\end{equation} 
Let us  write the equations of motion for these new amplitudes as 
the result of a combination of Eqs. \re{he1} and \re{lin} 
\begin{eqnarray}\label{heiz_alpha} 
\dot{\alpha}^{(\pm)}_i (\ve{p},t)&=&\frac12\Delta 
(\ve{p},t)\beta_i^{(\mp)}(-\ve{p},t) \pm i\omega(\ve{p},t) 
\alpha^{(\pm)}_i(\ve{p},t)+\eta_{ij}(\ve{p},t)\alpha_j(\ve{p},t),\nonumber\\ 
\dot{\beta}^{(\pm)}_i (-\ve{p},t)&=&\frac12\Delta 
(\ve{p},t)\alpha_i^{(\mp)}(\ve{p},t) \pm i\omega(\ve{p},t) 
\beta^{(\pm)}_i(-\ve{p},t)+\eta_{ij}(\ve{p},t)\beta_j(-\ve{p},t). 
\end{eqnarray} 
The spin rotation matrix $\eta_{ij}$ is defined as 
\begin{equation}\label{matrix1} 
\eta(\ve{p},t) = 
\begin{bmatrix} 
0 & \dot{\ve{e}}_1 \ve{e}_2 & \frac{\displaystyle 
\omega}{\displaystyle m}\dot{\ve{e}}_1\ve{e}_3  \\[5pt] -\dot{\ve{e}}_1 
\ve{e}_2& 0 & \frac{\displaystyle\omega}{\displaystyle m}\dot{\ve{e}}_2\ve{e}_3\\[5pt] 
-\frac{\displaystyle m}{\displaystyle 
\omega}\dot{\ve{e}}_1\ve{e}_3 & -\frac{\displaystyle 
m}{\displaystyle\omega}\dot{\ve{e}}_2\ve{e}_3& - \Delta_m, 
\end{bmatrix}. 
\end{equation} 
where $\Delta_m=-\dot{m}/m+\Delta$. Together with the Hamiltonian 
\re{hamp}, the operators of total momentum and charge take also 
a diagonal form. However, the spin operator 
\begin{equation}\label{spin} 
S_i=\varepsilon_{ijk}\int  d\ve{x}\,\left[ u^*_k \pi^*_j +\pi_j 
u_k - u^*_j \pi^*_k -\pi_k u_j\right] 
\end{equation} 
has a non-diagonal form in spin space in terms of the 
operators $\alpha^{(\pm)}$ and  $\beta^{(\pm)}$ 
\begin{equation}\label{spin1} 
S_k=i\varepsilon_{ijk}\sum\limits_{\ve{p}} 
\,\biggl[\alpha_i^{(+)}(\ve{p},t) \alpha_j^{(-)}(\ve{p},t)- 
\beta_i^{(-)}(-\ve{p},t)\beta_j^{(+)}(-\ve{p},t)\biggr]. 
\end{equation} 
In particular, the spin projection to the momentum $3$-axis is 
\begin{eqnarray}\label{s3} 
S_3&=&i\sum\limits_{\ve{p}}\, 
\biggl[\alpha_1^{(+)}(\ve{p},t)\alpha_2^{(-)}(\ve{p},t) 
-\alpha_2^{(+)}(\ve{p},t)\alpha_1^{(-)}(\ve{p},t)\nonumber\\ &+& 
\beta_2^{(-)}(-\ve{p},t)\beta_1^{(+)}(-\ve{p},t)- 
\beta_1^{(-)}(-\ve{p},t)\beta_2^{(+)}(-\ve{p},t) \biggr]. 
\end{eqnarray} 
Thus, this representation can be called an incomplete 
quasi-particle one with non-fixed spin projection. 
The operator \re{s3} can be diagonalized with a linear 
transformation to the circular polarized waves basis \cite{BS} 
\begin{eqnarray}\label{lin1} 
c_i^{(\pm)}(\ve{p},t) &=& R^{(\pm)}_{ik} 
\alpha_k^{(\pm)}(\ve{p},t),\nonumber\\ 
d_i^{(\pm)}(-\ve{p},t) &=& R^{(\pm)*}_{ik} 
\beta_k^{(\pm)}(-\ve{p},t), 
\end{eqnarray} 
with the unitary matrix 
\begin{equation}\label{matr2} 
R^\pm = \frac{1}{\sqrt{2}}\left[\begin{array}{ccc} 
  1 & \mp i& 0 \\ 
  \pm i & 1 & 0 \\ 
  0 & 0 & \sqrt{2} \\ 
\end{array}\right]. 
\end{equation} 
As a result, the new amplitudes $c^{(\pm)},d^{(\pm)}$ in the QPR 
correspond to the creation and annihilation operators of charged 
vector quasiparticles with the total 
energy, 3-momentum, charge and spin projection on the chosen 
direction 
\begin{eqnarray}\label{hamq} 
 H(t)&=&\sum\limits_{\ve{p}}\,\omega(\ve{p},t)\biggl[c_i^{(+)}(\ve{p},t) 
 c_i^{(-)}(\ve{p},t) 
  +d_i^{(-)}(-\ve{p},t)d_i^{(+)}(-\ve{p},t)\biggr], \\ 
\ve{\Pi}(t)&=& \sum\limits_{\ve{p}} 
 \,\ve{P}\biggl[c_i^{(+)}(\ve{p},t) 
   c_i^{(-)}(\ve{p},t) 
  - d_i^{(-)}(-\ve{p},t)d_i^{(+)}(-\ve{p},t)\biggr],\\ 
Q&=& e\sum\limits_{\ve{p}}\,\biggl[c_i^{(+)}(\ve{p},t) 
   c_i^{(-)}(\ve{p},t) 
-d_i^{(-)}(-\ve{p},t)d_i^{(+)}(-\ve{p},t)\biggr],\\ 
S_3(t)&=&\sum\limits_{\ve{p}}\, \biggl[c_1^{(+)}(\ve{p},t)c_1 
^{(-)}(\ve{p},t) - d_1^{(-)}(-\ve{p},t)d_1^{(+)}(-\ve{p},t) \nonumber\\ 
&+& d_2^{(-)}(-\ve{p},t)d_2^{(+)}(-\ve{p},t)- 
c_2^{(+)}(\ve{p},t)c_2^{(-)}(\ve{p},t) 
 \biggr]. 
\end{eqnarray} 
This representation can be named the complete 
quasi-particle one. The equations of motion for these amplitudes 
follow from Eqs. \re{heiz_alpha},\re{lin1} 
\begin{eqnarray}\label{heisc} 
\dot{c}^{(\pm)}_i (\ve{p},t)&=&\frac12\Delta 
(\ve{p},t)d_i^{(\mp)}(-\ve{p},t) \pm i\omega(\ve{p},t) 
c^{(\pm)}_i(\ve{p},t)+g_{ij}^{(\pm)}(\ve{p},t)c^{(\pm)}_j(\ve{p},t),\nonumber\\ 
\dot{d}^{(\pm)}_i (-\ve{p},t)&=&\frac12\Delta 
(\ve{p},t)c_i^{(\mp)}(\ve{p},t) \pm i\omega(\ve{p},t) 
d^{(\pm)}_i(-\ve{p},t)+\stackrel{*}{g}^{(\pm)}_{ij}(\ve{p},t)d^{(\pm)}_j(-\ve{p},t). 
\end{eqnarray} 
The the matrix $g_{ij}$ is defined as 
\begin{equation}\label{matrix2} 
g^{(\pm)}= 
\begin{bmatrix} 
\pm i\dot{\ve{e}}_1 \ve{e}_2 & 0 &\frac{\displaystyle 
\omega}{\displaystyle 
m}\dot{\ve{e}}^{(\mp)} \ve{e}_3  \\[5pt] 
0& \mp i\dot{\ve{e}}_1\ve{e}_2 & \frac{\displaystyle 
\omega}{\displaystyle 
m}\dot{\ve{e}}^{(\pm)}\ve{e}_3\\[5pt] 
-\frac{\displaystyle 
m}{\displaystyle\omega}\dot{\ve{e}}^{(\mp)}\ve{e}_3& 
-\frac{\displaystyle m}{\displaystyle\omega}\dot{\ve{e}}^{(\pm)} 
\ve{e}_3& - \Delta_m 
\end{bmatrix}, 
\end{equation} 
where $\ve{e}^{(\pm)} = (\ve{e}_1\pm i\ve{e}_2)/\sqrt{2}$. 
 
The transition to this representation from the initial one $(a,b)$ 
is defined by the combination of the transformations \re{lin} and 
\re{lin1}, 
\begin{eqnarray}\label{lin3} 
\ve{c}^{(\pm)}(\ve{p},t) &=& U^{(\pm)}(\ve{p},t) 
\ve{a}^{(\pm)}(\ve{p},t),\nonumber \\ 
\ve{d}^{(\pm)}(-\ve{p},t) &=& U^{(\pm)*}(\ve{p},t) 
\ve{b}^{(\pm)}(-\ve{p},t), 
\end{eqnarray} 
with non-unitary operator 
\begin{equation}\label{lin_u} 
U^{(\pm)}(\ve{p},t) = R^{(\pm)}\cdot E^{-1}(\ve{p},t) = 
\begin{bmatrix} 
  e_1^{(\mp)} & e_2^{(\mp)} & e_3^{(\mp)} \\ 
  e_1^{(\pm)} & e_2^{(\pm)} & e_3^{(\pm)} \\ 
  \frac{m}{\omega}e_{31} &  \frac{m}{\omega}e_{32} & 
   \frac{m}{\omega}e_{33} 
\end{bmatrix}. 
\end{equation} 
 
The quantization problem is to be solved while taking into account 
the equation of motion \re{heisc}. It leads to the following 
non-canonical commutation relations 
\begin{eqnarray}\label{cr} 
 \bigl[c^{\,(-)}_i(\ve{p},t),c^{\,(+)}_j(\ve{p'},t)\bigr]= 
 \bigl[d^{\,(-)}_j(\ve{p},t),d^{\,(+)}_i(\ve{p'},t)\bigr]= 
 Q_{ik}^{(-)}(\ve{p},t)Q_{jk}^{(+)}(\ve{p},t) \delta_{\ve{p}\ve{p}'}\,, 
\end{eqnarray} 
where the matrices $Q_{il}^{(\pm)}(\ve{p},t)$ are defined by the 
equations 
\begin{equation}\label{qq} 
 \dot{Q}_{ij}^{(\pm)}(\ve{p},t) = 
 g^{(\pm)}_{ik}(\ve{p},t)Q_{kj}^{(\pm)}(\ve{p},t) 
\end{equation} 
with the initial conditions 
\begin{equation}\label{qic} 
 \lim_{t\rightarrow -\infty}{Q}_{ij}^{(\pm)}(\ve{p},t) = 
 \delta_{ij}, 
\end{equation} 
i.e. the commutation relations \re{cr} transform to the 
canonical form only in the asymptotic limit $t\rightarrow 
-\infty$. The commutation relations \re{cr} provide the 
definition of positive energy quasiparticle excitations 
 with some time dependent energy reservoir of the vacuum.

\section{Kinetic equation}\label{KE} 
 
The standard procedure to derive the KE \cite{Smol} is based 
on the Heisenberg-type equations of motion \re{he1} or \re{heisc}. 
Let us introduce the one-particle correlation functions of vector 
particles and antiparticles in the initial ($a,b$)-representation 
\begin{eqnarray}\label{distr} 
F_{\mu\nu}(\ve{p},t)&=& 
 \langle 0_{in}|a^{(+)}_\mu (\ve{p},t) a^{(-)}_\nu 
 (\ve{p},t)|0_{in}\rangle ,\nonumber\\ 
\tilde{F}_{\mu\nu}(\ve{p},t) &=& 
 \langle 0_{in}|b^{(-)}_\mu (-\ve{p},t) b^{(+)}_\nu(-\ve{p},t) 
 |0_{in}\rangle , 
 \end{eqnarray} 
where the averaging procedure is performed over the in-vacuum state 
\cite{Grib94}. Differentiating the first one with respect to time, 
we obtain 
\begin{equation}\label{36} 
  \dot{F}_{\mu\nu}(\ve{p},t) =\frac12 \Delta (\ve{p},t) \left\{ 
  F^{(+)}_{\mu\nu}(\ve{p},t) + F^{(-)}_{\mu\nu}(\ve{p},t) \right\} , 
\end{equation} 
where the auxiliary correlation functions are introduced as 
\begin{equation}\label{anom} 
F_{\mu\nu}^{(\pm)}(\ve{p},t)= 
 \langle 0_{in}|a^{(\pm)}_\mu (\pm\ve{p},t) b^{(\pm)}_\nu 
(\pm\ve{p},t)|0_{in}\rangle . 
\end{equation} 
The equations of motion for these functions can be obtained by 
analogy with Eq. \re{36}. We write them out in the integral 
form \begin{equation}\label{38} 
 F^{(\pm)}_{\mu\nu}(\ve{p},t) =\frac12 \int\limits_{-\infty}^{t} 
  dt' \Delta(\ve{p},t') 
 \left[F_{\mu\nu}(\ve{p},t')+\tilde{F}_{\mu\nu}(\ve{p},t') 
 \right] e^{\mp 2i\theta(\ve{p};t,t')}, 
\end{equation} 
where 
\begin{equation}\label{39} 
  \theta(\ve{p}; t,t_0) = \int\limits_{t_0}^t dt' \omega(\ve{p},t'). 
\end{equation} 
In Eq. \re{38}, the asymptotic condition 
 $F^{(\pm)}_{\mu\nu}(\ve{p},-\infty) = 0$ 
(the absence of quasi-particles in the initial time) has been introduced. 
The substitution of Eq. \re{38} into Eq. \re{36} leads to the 
resulting KE 
\begin{equation}\label{ke} 
\dot{F}_{\mu\nu}(\ve{p},t)= \frac{1}{2} \Delta ( \ve{p},t) 
 \int\limits_{-\infty}^{t} dt' \Delta (\ve{p},t') [ 
F_{\mu\nu}(\ve{p},t')+\tilde{F}_{\mu\nu}(\ve{p},t') ] 
 \cos[ 2\, \theta (\ve{p};t,t') ]. 
\end{equation} 
This KE is an almost natural generalization of the corresponding KE 
for scalar particles \cite{Smol}. 
 
Thus, the OR turns out to be an effective method for the 
diagonalization of the Hamiltonian in the Fock space. It is 
sufficient for the derivation of the KE \re{ke}. However, 
on this stage, there is a number of problems that are 
specific for the vector field theory: the energy is not 
positively defined, the spin operator has a nondiagonal 
form in the space of spin states etc., see above. 
This circumstance hampers the physical interpretation of 
the distribution function \re{distr}. In order to overcome 
this difficulty, it is necessary to pass on to the complete 
QPR in which the system has well-defined values of energy, 
spin etc. The simplest way of derivation of the KE is the 
QPR, based on the application of the transformations 
\re{lin3} directly onto the KE \re{ke}.

\subsection{Kinetic equation in QPR}\label{iqpr} 
 
By analogy with the definitions \re{distr}, let us introduce the 
correlation  functions of vector particles and antiparticles in 
the complete QPR 
\begin{eqnarray}\label{distr1} 
  f_{ik}(\ve{p},t)&=& 
 \langle 0_{in}|c^{(+)}_i (\ve{p},t) c^{(-)}_k (\ve{p},t)|0_{in}\rangle , 
 \nonumber\\ \tilde{f}_{ik}(\ve{p},t) &=& 
 \langle 0_{in}|d^{(-)}_i (-\ve{p},t) d^{(+)}_k(-\ve{p},t)|0_{in}\rangle . 
\end{eqnarray} 
They are connected with the primordial correlation functions 
\re{distr} by relations of the type 
\begin{equation}\label{dfdf} 
f_{ik}(\ve{p},t)=U^{+}_{in}(\ve{p},t)U^{-}_{km}(\ve{p},t) 
F_{nm}(\ve{p},t), 
\end{equation} 
where $F_{nm}(\ve{p},t)$ is the "spatial" part of the tensor 
function $F_{\mu\nu}(\ve{p},t)$ \re{distr} ($m,n=1,2,3$). 
 
To obtain the resulting KE in the complete QPR we differentiate 
the function $f_{ik}(\ve{p},t)$ \re{dfdf} with respect to time, 
and take into account the KE \re{ke} 
\begin{multline}\label{keq} 
\dot{f}_{ik}(t)=  \dot{U}^{(+)}_{ij}(t) U^{(+)-1}_{km}(t) 
f_{jm}(t) + \dot{U}^{{(-)}}_{kj}(t) 
U^{(-)-1}_{jm}(t) f_{im}(t) +\\ 
+\frac12 U^{(+)}_{ij}(t) U^{(-)}_{kl}(t) 
\Delta(t)\int\limits^t_{-\infty} dt' \Delta(t') 
U^{(+)-1}_{jm}(t')U^{(-)-1}_{ln}(t') [ f_{mn}(t') + 
\tilde{f}_{mn}(t') ] \cos{2\theta(\ve{p};t,t')}. 
\end{multline} 
In comparison with non-Markovian effects of vacuum 
tunneling of scalar particles \cite{Schmidt99}, the 
considered case has its own characteristics related to the 
dynamics of spin twist. The Markovian limit ($f_{mn} (t') 
\rightarrow f_{mn}(t)$ in the integral part of r.h.s. of 
\re{keq}) is  admitted for rather slow processes and 
results in a significant simplification of the KE \re{keq}. 
 
The system of the integro-differential Eqs. \re{keq} can be reduced to 
a system of 27 coupled ordinary differential equations that is 
convenient for numerical calculations. We will not analyze here this 
rather complicated case and will restrict ourselves below to the 
consideration of a simple particular case having cosmological 
motivation. 
 
\subsection{Deformation of the energy gap} 
 
Let us consider the vacuum creation of vector bosons in the 
case when it is caused by an arbitrary time dependent 
deformation of energy gap, i.e.  $m=m(t)$ and $A^k(t)=0$. 
This is an isotropic case with $P_k=p_k$ and 
$\dot{\ve{e}}_i=0$. These conditions are rather 
characteristic for different inflationary models of the 
preheating process (see, e.g., \cite{18} and the references 
given there). As a result, the KE \re{keq} takes the 
following form
\begin{multline}\label{keq1} 
\dot{f}_{ik}(\ve{p},t)= -\Delta_m(\ve{p},t) [ 
\delta_{i3}f_{3k}(\ve{p},t) + \delta_{k3}f_{i3}(\ve{p},t) ]+\\ 
+\frac12 \Delta(\ve{p},t)\int\limits^t_{-\infty} dt' 
\Delta(\ve{p},t')  M_{ikjl}(\ve{p},t,t') [ f_{jl}(\ve{p},t') + 
\tilde{f}_{jl}(\ve{p},t') ] \cos{2\theta(\ve{p};t,t')}, 
\end{multline} 
where \begin{equation} 
 M_{ikjl}(t,t') = \delta_{ij}^\perp \delta_{kl}^\perp + 
 \frac{\omega(t')}{\omega(t)}\frac{m(t)}{m(t')}\left[ 
\delta_{i3}\delta_{j3} \delta^\perp_{kl} + \delta_{k3}\delta_{l3} 
\delta^\perp_{ij} + \frac{\omega(t')}{\omega(t)}\frac{m(t)}{m(t')} 
 \delta_{i3}\delta_{k3}\delta_{j3}\delta_{l3}\right] 
\end{equation} 
and $\delta_{ik}^\perp =\delta_{ik}-\delta_{i3}\delta_{k3}$. 
 
As to be expected, the distribution functions 
$f_{\alpha\beta}(\ve{p},t)$ and $F_{\alpha\beta}(\ve{p},t)$ 
satisfy the same KE \re{ke} for $\alpha=1,2$. The feature of the 
complete QPR becomes apparent only in the component of tensor 
distribution function $f_{ik}(\ve{p},t)$ that contains the 
preferred values of spin index $i,k=3$. Let us select the KE for 
the diagonal components of the correlation functions \re{distr1} 
having a direct physical meaning as the distribution functions of the 
transversal ($i=1,2$) and longitudinal components
\begin{eqnarray} 
  \dot{f}_{i}(\ve{p},t) &=& \frac12 \Delta(\ve{p},t) 
  \int\limits^t_{-\infty} dt' \Delta(\ve{p},t') 
\bigl[1+ 2 f_{i}(\ve{p},t')\bigr] 
\cos{2\theta(\ve{p};t,t')}, \quad \label{kema}\\ 
\dot{f}_{3}(\ve{p},t) &=& -2\Delta_m(\ve{p},t) 
f_{3}(\ve{p},t)+\nonumber\\ 
&+&\frac12 \Delta(\ve{p},t) \frac{m^2(t)}{\omega^2(t)} 
\int\limits^t_{-\infty} dt' \Delta(\ve{p},t') 
\frac{\omega^2(t')}{m^2(t')} \bigl[2f_{3}(\ve{p},t')+ 
Q(\ve{p},t')\bigr] \cos{2\theta(\ve{p};t,t')}. \label{kem3} 
\end{eqnarray} 
Here the shorthand notation $f_{ii}=f_i$ has been introduced 
for the diagonal components of the matrix 
correlation functions \re{distr1}, and we have 
\begin{eqnarray}\label{dd} 
\Delta =\frac{m \dot{m}}{\omega^2}, \qquad \Delta_m = -\Delta 
\frac{\ve{p}^2}{m^2}. 
\end{eqnarray} 
It is possible to show, that  the distribution functions of the 
longitudinal ($i=3$) and transversal ($i=1,2$) components are 
connected by the relation 
\begin{equation}\label{48} 
    f_{3}(\ve{p},t) = Q (\ve{p},t)f_{1}(\ve{p},t), 
\end{equation} 
where  $Q (\ve{p},t)$ is the function occurring in the 
commutator of the creation and annihilation operators for 
the longitudinal bosons, 
\begin{eqnarray}\label{49} 
[c_3^{(-)}(\ve{p},t), c_3^{(+)}(\ve{p'},t)]&=& Q(\ve{p},t)\, \delta_ 
{pp'},  \nonumber\\[6pt] 
 Q (\ve{p},t) &=& \exp{\biggl[-2\int_{t_0}^t \Delta_m(t') dt'\biggr]} = 
 \left[\frac{m(t)}{m(t_0)}\frac{\omega(t_0)}{\omega(t)}\right]^2, 
\end{eqnarray} 
where $m(t_0)$ and $\omega (t_0)$ are some initial values 
of $m(t)$ and $\omega (t)$.

Owing to Eq. \re{48}, it is sufficient to solve the one 
equation \re{kema}. We use now the well-known procedure of 
the reduction of the KE from the integro-differential form 
to the corresponding system of ordinary differential 
equations \cite{Vinnik01} in order to study the KE 
\re{kem3} numerically and to investigate the asymptotic 
behavior of its solutions for large momenta below in  
Sect. \ref{EOS}, 
\begin{equation}\label{58} \dot{f}_k=\frac12\Delta u_k,\qquad 
\dot{u}_k = \Delta (1 + 2{f}_k) - 2\omega v_k,\qquad 
\dot{v}_k =2\omega u_k . 
\end{equation} 
Here  $u_{k}$ and $v_{k}$ ($k=1,2$) are some auxiliary 
functions responsible for the different effects of vacuum 
polarization (see, e.g., \cite{Vinnik01}). 
It can be shown by analogy with the scalar field case \cite{OR} 
that this system is conservative and has the first integral of 
motion 
\begin{equation}\label{59} 
    (1+2f_k)^2 - u_k^2 -v_k^2 = 1, 
\end{equation} 
if the particles are absent in the initial time, 
\begin{equation}\label{60} 
    f_k(-\infty)=u_k(-\infty)=v_k(-\infty)=0. 
\end{equation} 
The general initial condition for all diagonal components 
of the distribution function 
\begin{equation}\label{ic} 
  \lim_{t \rightarrow - \infty} f_k(t) = 
  \lim_{t \rightarrow - \infty} u_k(t) = 
  \lim_{t \rightarrow - \infty} v_k(t) = 0 
\end{equation} 
leads to the following requirement 
\begin{equation}\label{icm} 
  \lim_{t \rightarrow - \infty} m(t) = m_0,  \quad \mbox{or}\quad 
  \lim_{t \rightarrow - \infty} \dot{m}(t) = 0. 
\end{equation} 

The main characteristic of the vacuum creation process is 
the total number density of vector  bosons, 
\begin{equation}\label{dens} 
  n_{tot}(t) = 2\sum\limits^3_{i=1} n_i(t)= \frac{1}{\pi^2} 
 \int\limits^\infty_0 p^2 dp\, \bigl[2f_1(p,t) 
+f_3(p,t)\bigr]= \frac{1}{\pi^2} \int\limits^\infty_0 p^2 
dp\, f_1(p,t) \bigl[ 2+Q(p,t)\bigr], 
\end{equation} 
where isotropy of the system was taken into account, 
$p=|\ve{p}|$. The factor $2$ corresponds to equal numbers of 
particles  and anti-particles. As it will be shown in 
Sect. \ref{EOS}, the integral \re{dens} is convergent. 
 
Up to now, the time dependence of particle mass was brought 
in on the phenomenological level, without an indication of 
any concrete mechanism of its origin. As it was shown 
above, this is sufficient for the formal construction of kinetic 
theory of vacuum particle creation in the case of rather 
fast mass change. We will consider below the case when 
the time dependence of vector boson mass is defined by the 
conformal evolution of the universe.

\section{Vector boson production in the early Universe}
\label{UNI} 
 
The description of the vacuum creation of particles in the time 
dependent gravitational fields of cosmological models goes back to
Refs. \cite{Parker,GM,SU,Z} and has been recently reviewed, e.g., 
in the monographs \cite{Birrel,ZN}. 
The specifics of our work consits in the consideration of vacuum 
generation of vector bosons in the conditions of the early 
Universe in the framework of a conformal-invariant 
cosmological model \cite{Perv02}, thus assuming, that the 
space-time is conformally flat and that the expansion of 
the Universe in the Einstein frame (with metric 
$\tilde{g}_{\mu\nu}$) with constant masses $\tilde{m}$ 
can be replaced by the change of 
masses in the Jordan frame (with metric $g_{\mu\nu}$) due 
to the evolution of the cosmological (scalar) dilaton 
background field \cite{DB,FM}. This mass change is defined 
by the conformal factor $\Omega(x)$ of the conformal 
transformation 
\begin{equation} 
\tilde{g}_{\mu\nu}(x) = \Omega^2 (x)g_{\mu\nu}. 
\end{equation} 
As mass terms generally violatate conformal invariance a
space-time dependent mass term 
\begin{equation} 
{m}(x) = \frac{1}{\Omega(x)}\tilde{m} 
\end{equation} 
has been introduced which formally keeps the conformal invariance 
of the theory \cite{BD}. 
In the important particular case of the isotropic FRW space-time, 
the conformal factor is equal to the scale factor, $\Omega(x)= 
a(\tilde{t})$, and hence $m(\tilde{t}) = a(\tilde{t})m_{obs}$, 
where $\tilde{t}$ is "the Einstein time" and $m_{obs}$ is the 
observable present-day mass. Such a dependence was used, e.g., in 
Ref. \cite{Grib94} for the Robertson-Walker metric. 
On the other hand, the scaling factor $a(\tilde{t})$ is defined by 
the cosmic equation of state (EoS). 
For a barotropic fluid, this EoS has the form 
\begin{equation} 
p_{ph} = (\gamma - 1)\epsilon_{ph} = c^2_s \epsilon_{ph}, 
\end{equation} 
where $p_{ph}$ and $\epsilon_{ph}$ are phenomenological pressure 
and energy density (in the distinction from "dynamical" $p$ and 
$\epsilon$ (see Sect. \ref{EOS}) below),  $\gamma$ is the 
barotropic parameter, $c_s$ is the sound velocity. The solution of 
the Friedman equation for such EoS leads to the following scaling 
factor
\begin{equation}\label{scaling} 
a(\tilde{t}) \sim \tilde{t}^{\ 2/3\gamma}. 
\end{equation} 
The kinetics of vacuum creation of massive vector bosons (Sects. II 
and III) was constructed in the flat Jordan frame with the proper 
conformal time ${t}$, which is necessary to introduce now in the 
Eq. \re{scaling}. The transition to the conformal time is defined 
by relation $dt = d\tilde{t}/a(\tilde{t})$. From this relation and 
Eq. \re{scaling} follows 
\begin{equation} 
\tilde{t} \sim \left[\left(1 - \frac{2}{3\gamma}\right)\, 
t\right]^{3\gamma / (3\gamma - 2)}. 
\end{equation} 
The substitution of this relation in the Eq. \re{scaling} leads to 
the mass evolution law in the terms of the conformal time
\begin{equation}\label{mass} 
m(t) = (t/t_H ) ^{\alpha}\, m_W, \qquad  \alpha = 
\frac{2}{3\gamma - 2}, 
\end{equation} 
where $t_H = [(1+\alpha)H]^{-1}$ is the scaling factor (the 
age of the Universe), $H$ is the Hubble constant and $m_W = 
80$GeV is the W-boson mass. Let us write here the values of 
the parameter $\alpha$ for some popular EOS: $\gamma =2$, 
$\alpha = 1/2 $ (stiff fluids); $\gamma = 4/3$, $\alpha = 
1$ (radiation); $\gamma =1$, $\alpha = 2$ (dust).

\begin{figure}[ht] 
\centering 
\includegraphics[width=75mm,height=60mm]{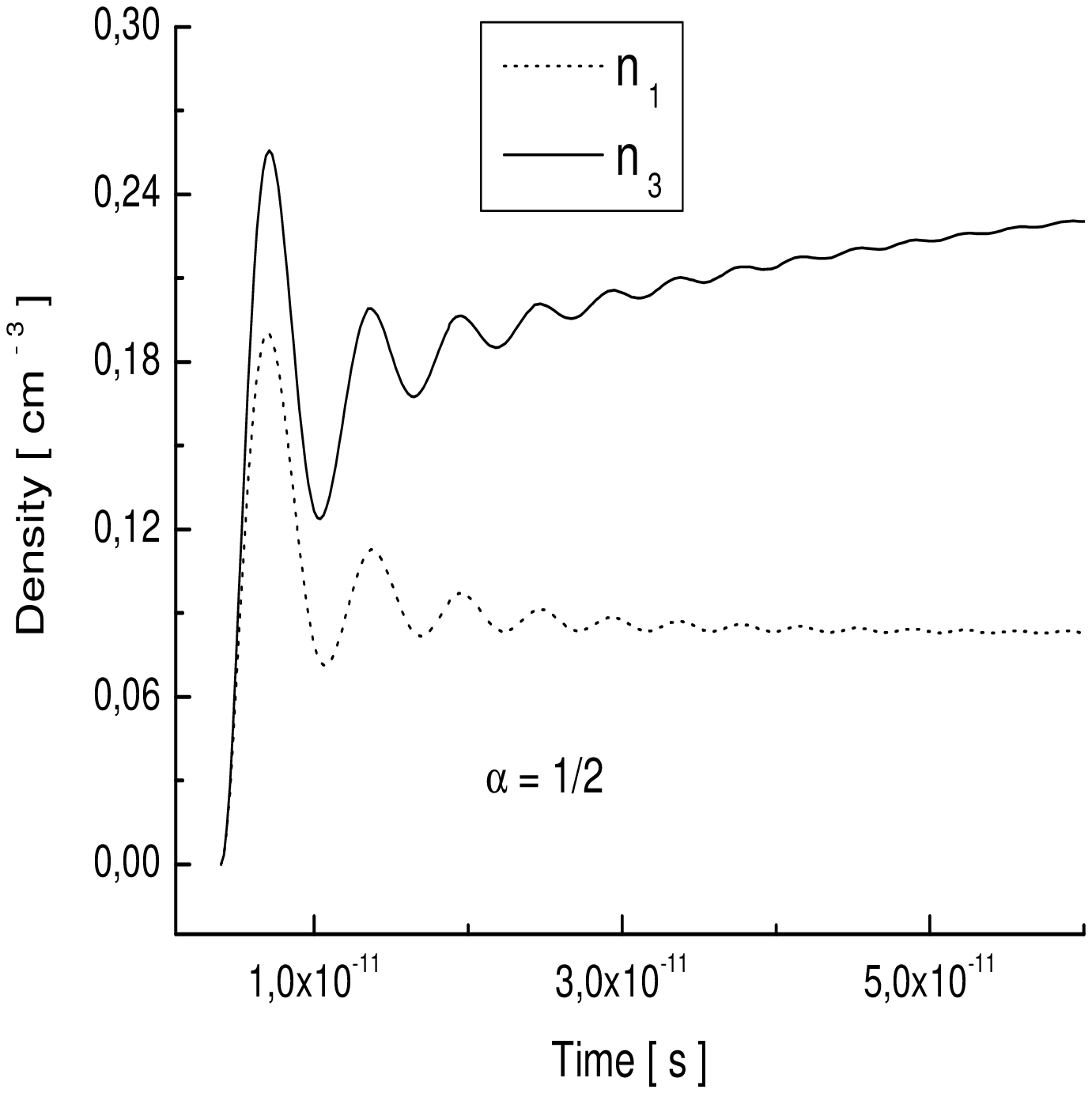}\hspace{10mm} 
\includegraphics[width=75mm,height=60mm]{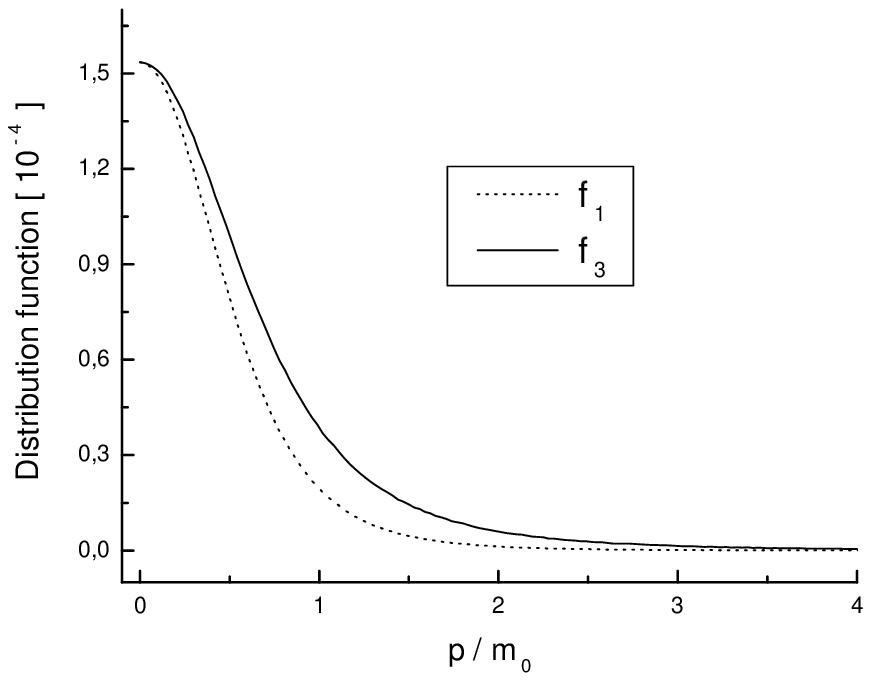} 
\caption{Time evolution of the particle number density with 
initial condition $m_0 \cdot t_0 = 1 $ for $\alpha=1/2$ 
(left) and the corresponding momentum distribution at the 
time $t \gg t_0$ (right).}\label{time} 
\end{figure}

\begin{figure}[ht] 
\centering 
\includegraphics[width=75mm,height=60mm]{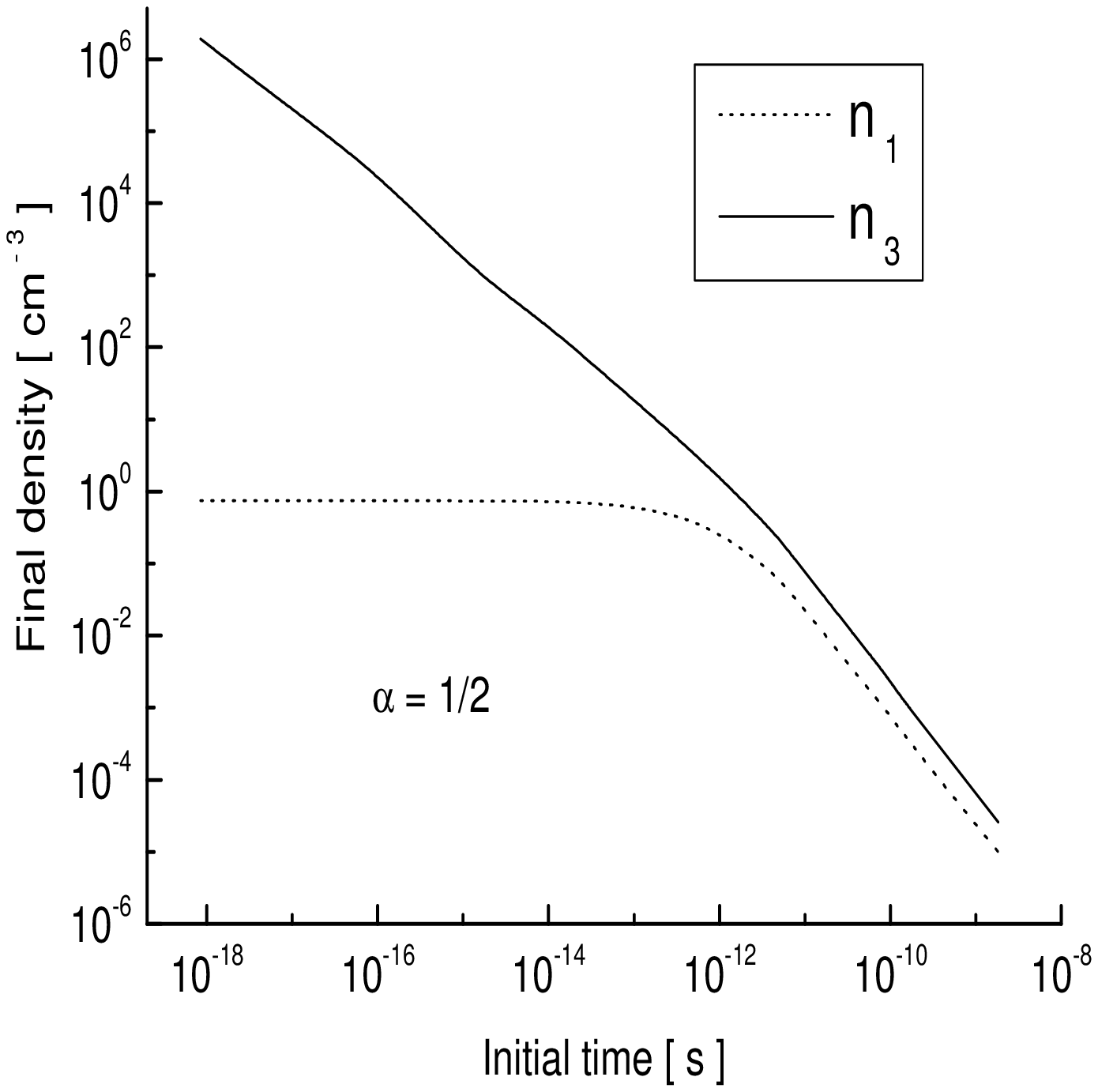}\hspace{10mm} 
\includegraphics[width=75mm,height=60mm]{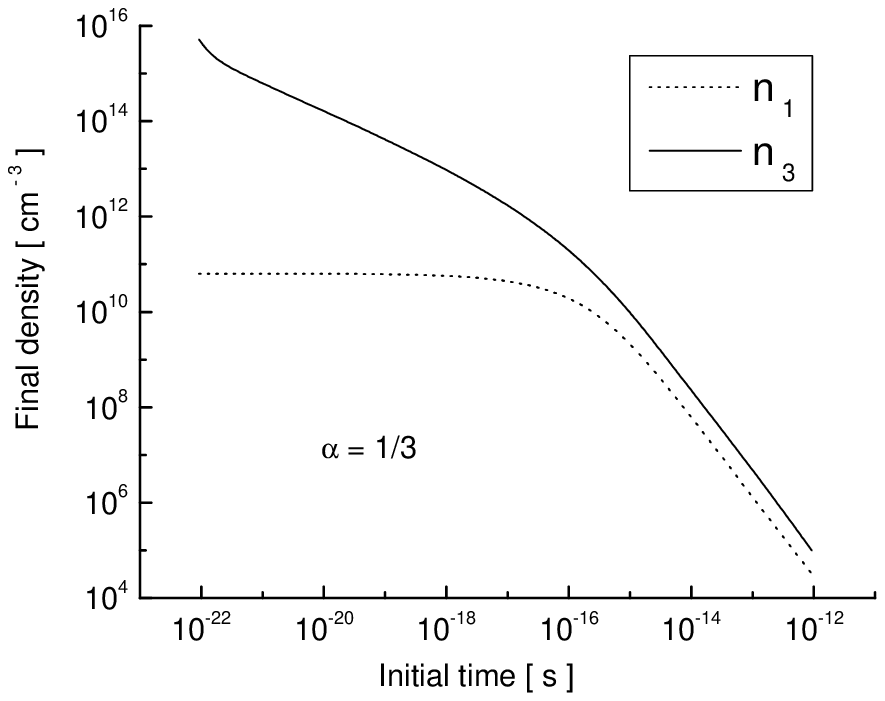} 
\caption{The dependence of the final particle number 
density on the initial time:  $\alpha=1/2$ (left), 
$\alpha=1/3$ (right).}\label{final} 
\end{figure} 

Due to back reactions and dynamical mass generation during 
the cosmic evolution the detailed mass history remains to 
be worked out. The central question, however, is whether 
the number density of produced W-bosons could be of the 
same order as that of the CMB photons, $n_{\rm CMB}\sim 465
~{\rm cm}^{-3}$. If this question could be answered 
positively, the vacuum pair creation of W-bosons from a 
time-dependent scalar field (mass term) could be suggested 
as a mechanism for the generation of matter and radiation 
in the early Universe. The non-Abelian nature of the 
W-bosons could even imply consequences for the generation 
of the baryon (and lepton) asymmetry due to topological 
effects \cite{DB}. 
 
The numerical analysis of the Eqs. \re{58} is performed by 
a standard Runge-Kutta method on a one-dimensional momentum 
grid. As one can see from Fig. 1, the  creation process 
ends very quickly and the particle density saturates at 
some final value. 
The momentum distribution of particles is formed also very 
early when $m(t)\approx m_0$ and frozen in such 
form so that later on for times $t \gg t_0$ most of the 
particles have very small momentum $p \ll m(t)$. The 
spectrum of created  bosons is essentially non-equilibrium, 
hence we should continue further the analysis of relevant 
dissipative mechanisms and other 
observable manifestations of the non-equilibrium state
(e.g., CMB photons; in this 
connection, see, for example, \cite{Komatsu}). 
 
The other interesting feature of the mechanism of 
particle production considered above is the qualitatively
different form of momentum distribution when compared to 
the creation in the electric field. 
As is known, the electric field generates the pairs 
with monotonic momentum distribution and $f(p=0)=max$ in 
the fermion case and with non-monotonic distribution and 
$f(p=0)=0$ in the boson case \cite{Grib94}. 
The creation mechanism connected with time-dependent 
effective mass modifies this situation to the contrary: 
the boson have a monotonic distribution (Fig. \ref{time}) 
and the fermions a non-monotonic one. 
The dependence of the corresponding final value of density on 
the initial time is shown in Fig. \ref{final}. The final 
density $n_1$ of particles with spin projection $\pm 1$ 
reaches a maximum, when we let the initial time go to very 
early times, close to the birth of the Universe. 
However, in the same limit, the density $f_3$ of particles 
with spin projection zero grows beyond all bounds. 
The choice of the EoS changes drastically the quantity of 
the created particles, thus giving values which are too small 
($\alpha = 1/2$) or too large ($\alpha = 1/3$) in comparison with 
the observed CMB photon densities. In order to improve this 
model, we should use an improved EoS, assuming that the 
barotropic parameter $\gamma$ 
charcterizing the evolution of the particle masses can change 
during the time evolution. Such a time-dependence could be induced 
by the back-reaction of the created particles on the scalar field. 
Furthermore, we could use another space-time model, e.g., 
the Kasner space-time \cite{SmolHP} instead of the conformally 
flat de Sitter one. The main achievement relative to the 
earlier work \cite{DB} is that in the present approach, 
there is no divergence in the distribution function, thus 
we do not need to introduce some ambiguous regularization 
procedure. 
 
\section{EOS for the isotropic case}\label{EOS} 
 
The relations for the energy density and pressure can be 
obtained from the energy-momentum tensor corresponding to 
the Lagrangian \re{eq2} 
\begin{equation}\label{e1} 
  T_{\mu\nu} = -\partial_{\mu}u^*_{\alpha} \partial_{\nu}u^{\alpha} 
  -\partial_{\nu}u^*_{\alpha} \partial_{\mu}u^{\alpha} - 
  g_{\mu\nu}\mathcal{L}. 
\end{equation} 
Thus, we have $\varepsilon = \langle 0_{in}| T_{00}|0_{in} 
\rangle$ and $p = 1/3\langle 0_{in}| T_{ii}|0_{in} 
\rangle$. Let us substitute the decompositions \re{or} and 
use the relations, which are a consequence of the 
spatial homogeneity of the system (in the discrete momentum 
representation) 
\begin{eqnarray}\label{e2} 
   \langle 0_{in}| a^{(+)}_{\mu}(\ve{p},t) a^{(-)}_{\nu}(\ve{p}',t) 
   |0_{in} 
\rangle &=& (2\pi)^3 \delta_{\ve{p},\ve{p}'} F_{\mu\nu} 
   (\ve{p},t), \nonumber \\ 
   \langle 0_{in}| b^{(-)}_{\mu}(-\ve{p},t) b^{(+)}_{\nu}(-\ve{p}',t) 
   |0_{in} 
\rangle &=& (2\pi)^3 \delta_{\ve{p},\ve{p}'} 
\tilde{F}_{\mu\nu} 
   (\ve{p},t), \\ 
   \langle 0_{in}| a^{(\pm)}_{\mu}(\ve{p},t) b^{(\pm)}_{\nu}(-\ve{p}',t) 
   |0_{in} 
\rangle &=& (2\pi)^3 \delta_{\ve{p},\ve{p}'} 
F^{(\pm)}_{\mu\nu} 
   (\ve{p},t). \nonumber 
\end{eqnarray} 
As a result we obtain the following expressions for the 
energy density and the pressure (after the transition to the 
thermodynamic limit, $d\,\Gamma = (2\pi)^{-3} d^3 p$) 
\begin{eqnarray} 
   \varepsilon(t) &=& -\int d\,\Gamma\, 
    \omega(\ve{p},t) [ F_{\mu}^{\mu} 
   (\ve{p},t) + \tilde{F}_{\mu}^{\mu} (\ve{p},t)], \label{e3a} \\ 
   p(t) &=& \frac{1}{3} \int \frac{d\,\Gamma}{\omega(\ve{p},t)} \biggl\{ 
   -\ve{p}^2[F_{\mu}^{\mu} (\ve{p},t) + \tilde{F}_{\mu}^{\mu} 
   (\ve{p},t)]  \nonumber \\ 
  &+& (2\ve{p}^2+3m^2)[F_{\mu}^{(+)\mu} (\ve{p},t) + F_{\mu}^{(-)\mu} 
(\ve{p},t)]   \biggr\}. \label{e3b} 
\end{eqnarray} 
We perform now the series of the consecutive 
transformations  of the functions 
$F_{\mu}^{\mu}(\ve{p},t)$, 
$\tilde{F}_{\mu}^{\mu}(\ve{p},t)$ and 
$F_{\mu}^{(\pm)\mu}(\ve{p},t)$: the exclusion of the 
$\mu=0$ component (according to Eq. \re{add2}) and the 
transition to the complete QPR with the help of Eqs. 
\re{lin_u}. Thus, we arrive at 
\begin{eqnarray}\label{e4} 
   F_{\mu}^{\mu}(\ve{p},t) = -\sum_{l=1}^{3} f_l(\ve{p},t), \;\; 
   \tilde{F}_{\mu}^{\mu}(\ve{p},t) = -\sum_{l=1}^{3} \tilde{f}_l(\ve{p},t), 
   \\ 
   F_{\mu}^{(\pm)\mu}(\ve{p},t) = -\sum_{l=1}^{3} f^{(\pm)}_l(\ve{p},t) 
    - 2\frac{p^2}{m^2}f_3^{(\pm)}(\ve{p},t), 
\end{eqnarray} 
where $f_l(\ve{p},t)$, $\tilde{f}_l(\ve{p},t)$ and 
$f^{(\pm)}(\ve{p},t)$ are the corresponding functions in 
the complete QPR. 
The substitution of these relations into Eqs. \re{e3a} and 
\re{e3b} leads to the following expressions for the energy 
density and pressure 
\begin{eqnarray}\label{e5} 
\varepsilon(t) &=& 2\sum_{l=1}^{3} \int d\,\Gamma 
\omega(\ve{p},t) 
f_l(\ve{p},t), \\ 
p(t) &=& \frac{1}{3}\sum_{l=1}^{3}\int 
\frac{d\,\Gamma}{\omega(\ve{p},t)}\biggl\{ 
2 p^2 f_l(\ve{p},t) - [2\omega^2(\ve{p},t)+m^2]  \nonumber \\ 
&&\times 
 \left( \frac{2 p^2}{3 m^2}\left[f_3^{(+)}(\ve{p},t)+ 
f_3^{(-)}(\ve{p},t)\right] + \left[ 
f_l^{(+)}(\ve{p},t)+f_l^{(-)}(\ve{p},t)\right]\right)\biggr\}. 
\end{eqnarray} 
Taking into account the isotropy of the system, we obtain 
the EOS for the massive vector boson gas 
\begin{align}\label{e6} 
    \varepsilon(t) &=& 2 \int d\,\Gamma \omega 
    ( 2 + Q ) f_1, \\ 
p(t) &=& \frac{2}{3} \int \frac{d\,\Gamma }{\omega} \left[ 
2 \ve{p}^2 (2+Q)f_1  \right] + \delta p_{vac} 
(t)\label{e7}, 
\end{align} 
where $\delta p_{vac} (t)$ is the contribution in the 
pressure induced by the vacuum polarization, 
\begin{equation}\label{e7a} 
\delta p_{vac} (t) = -  \frac{2}{3} \int \frac{d\,\Gamma 
}{\omega}\, \biggl(2\omega^2 + m^2 \biggr) \left[ 1+ Q 
\left(\frac{1}{2}+ \frac{\ve{p}^2}{m^2}\right)\right] u_1. 
\end{equation} 
In order to prove the convergence of the integrals \re{e6} - 
\re{e7a} we investigate the asymptotic behavior of the solution
to the system of Eqs. \re{58}. This system can be solved exactly 
in the asymptotic limit $p\gg m$ for the case $\alpha=1/2$ when
it gets the form 
\begin{eqnarray}\label{p1} 
\dot{f} &=& \frac{m_H^2}{4t_H}\frac{1}{ p^2} u,\nonumber \\ 
\dot{u} &=& \frac{m_H^2}{2t_H}\frac{1}{ p^2}(1+2f) - 2p v, \nonumber\\ 
\dot{v} &=& 2 p u~. 
\end{eqnarray} 
The solution of \re{p1} with the initial conditions \re{ic} is 
\begin{eqnarray}\label{p2} 
f(p,t)&=& \frac{\sin^2{p(t-t_0)}}{16 (p/m_0)^6}, \nonumber\\ 
u(p,t)&=& \frac12 v(p,t) = \frac{\sin{2 p(t-t_0)}}{4 
(p/m_0)^3}, 
\end{eqnarray} 
where $m_0=m(t_0)$. The numerical investigation of the 
general Eqs. \re{58} shows that the basic features of 
the solutions \re{p2} for $\alpha = 1/2$ are conserved also 
for other $\alpha > 0$. 
 
\begin{figure}[ht] 
\centering 
\includegraphics[width=75mm,height=60mm]{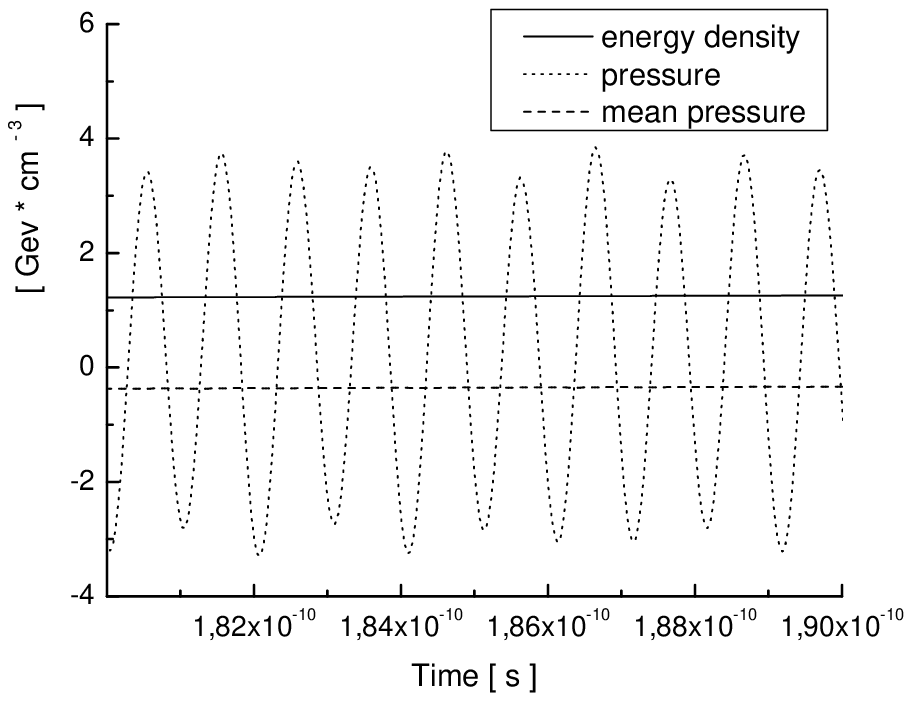}\hspace{10mm} 
\includegraphics[width=75mm,height=60mm]{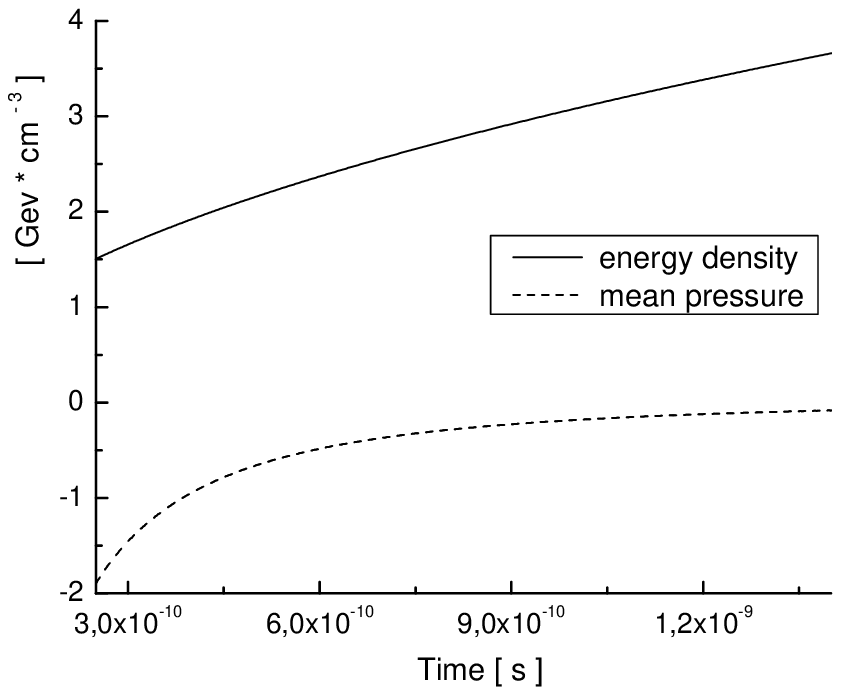} 
\caption{Time dependence of energy density and pressure at 
$t\gg t_0$ with initial condition $m_i\cdot t_i=1$ for 
$\alpha=1/2$.}\label{press} 
\end{figure}

According to \re{p2}, the particle and energy densities 
(Eqs. \re{dens} and \re{e6}, respectively) are 
convergent, but the vacuum polarization contribution to the 
pressure \re{e7a} is divergent. Moreover, irrelevant fast 
vacuum oscillations of the pressure are observed here. Let us 
remark that such a behavior of the pressure for a plasma created 
from the is not a special feature of the 
present theory but is characteristic also for the models where an 
electron - positron plasma is created in strong, time-dependent 
electrodynamic fields as investigated in \cite{23}. 
   The standard regularization procedure of similar 
integrals with some unknown functions satisfying
ordinary differential equations is based on the investigation of 
asymptotic decompositions of these functions in power series of 
the inverse momentum, $1/|\vec{p}|^n$ (the 
procedure of n-waves regularizations \cite{nw}). In the 
considered case, such a procedure is not effective because 
the solutions \re{p1} has fastly oscillating factors ("zitterbewegung"),
the asymptotic decompositions of which 
lead to the secular terms. Therefore we regularize the 
pressure by a momentum cut-off at $p=10m_0$ and separate 
its stable part by the time averaging 
\begin{equation}\label{3} 
<p> = \frac{1}{(t-t_0)}\int\limits_{t_0}^t p(t) dt. 
\end{equation} 
The such "coarse graining" procedure was proposed in 
\cite{19} in order to exclude the "zitterbewegung" from the
description of vacuum particle creation. In reality, the 
smoothing of these  fast oscillations occurs due to 
dissipative processes that are not taken into account here.

Fig. \ref{press} shows that the mean pressure remains 
negative and its magnitude becomes negligible in comparison 
with the energy density. At large times the energy 
density grows but the pressure stays very small, $p \simeq 0$. 
The energy growth with the condition $p \simeq 0$ leads the 
conclusion mentioned in Sect. \ref{UNI} that the
massive vector boson-anti-boson gas created from the vacuum 
is a cold one (see also Fig. 1).  
It can be seen directly from Eq. \re{e6}, that at large 
times $\epsilon (t) \simeq m(t)n(t)$, because of 
$\omega (t) \simeq m(t)$. Let us remark also 
that such an EoS of the massive vector boson gas
($\epsilon \neq 0$ and $p \simeq 0$) corresponds to dust-like 
matter \cite{20}, which would characterize the evolution of 
the Universe during those stages when the vector boson gas 
is the dominant component of its matter/energy content. 
On the qualitative level, this conclusion is valid independent 
of the concrete choice of an EoS and, in particular, 
in the case of the dust-like EoS. 
In a subsequent work, we plan to obtain a 
formula of the type \re{mass} as a result of the solution of 
the Friedman equation with the EoS \re{e6}-\re{e7a} 
(such a procedure represents the back reaction problem) 
and to investigate self-consistently the production of vector 
bosons in the Universe.

\section{Summary}\label{SUM} 
 
The present work was devoted to the kinetic description of 
vacuum creation of massive vector bosons caused either by 
a time dependence of the mass or by the action of a
non-stationary electric field. The statement of the problem 
is stimulated by modern cosmological problems related to 
the need for an explanation of the nature of the recently
uncovered accelerating expansion of the Universe. 
The resulting KE \re{keq} of 
non-Markovian type is obtained within a powerful non-perturbative 
framework and using the OR which provides an effective approach 
to the QPR, in the language of which the kinetic theory 
is constructed. We apply then this KE for the analysis of 
the important particular case of an isotropic gas of vector 
bosons with a time dependent mass which can be justified 
on the basis of a conformal-invariant scalar-tensor theory of
gravitation. We show that the kinetic theory leads 
to a reasonable density of vector bosons in an early period 
of the Universe evolution which is sufficient to
explanation the present density of CMB photons. 
 
The obtained results constitute a foundation for the 
subsequent investigation of the dynamics of vector bosons 
created from the vacuum (the equation of state, the long 
wave-length acoustic excitations, the back-reaction problem 
etc.). It is necessary to underline that we have considered 
here the single mechanism of mass change  induced by the 
conformal expansion of the Universe. It was necessary to 
switch on the mass $m_0 (t_0)$ at some arbitrary initial time 
$t_0$. Thus, for the construction of a more consistent theory, 
one should eventually take into 
account the  inflation mechanism  of mass generation acting 
during an earlier period the Universe evolution 
\cite{18,22,24,25}. 
 
It is very interesting to investigate the 
generation of elementary particles of different masses. 
Eq. \re{mass} is valid for all elementary  particles 
independent of their inner symmetry class. Then the 
chemical composition of created matter  and the EoS of the 
Universe must be fixed rather precisely and  can be subject to 
experimental verification.

\subsection*{Acknowledgement}
This work was partly supported by a Russian Federations 
State Committee for Higher Education grant No. E02-3.3-210 
and RFBR grant No. 03-02-16877. S.A.S. acknowledges  by DFG 
grant No. 436 RUS 117/78/04, D.B. was supported by the 
Virtual Institute of the Helmholtz association under grant No. 
VH-VI-041. We are grateful to Prof. M.P. D\c{a}browski for 
discussion of some cosmological aspects of the present work.


\begin{thebibliography}{99} 
 
\bibitem{Nik} A.I. Nikishov, JETP \textbf{93}, 197 (2001). 
 
\bibitem{Grib94} A.A. Grib, S.G. Mamaev and V.M. Mostepanenko, 
{\it Vacuum Quantum Effects in Strong External Fields}, 
 (Friedmann Laboratory Publishing, St. Peterburg, 1994). 
 
\bibitem{Pavel} H.-P. Pavel and V.N. Pervushin, Int. J. Mod. Phys. 
A \textbf{14}, 2285 (1999). 
 
\bibitem{VT} V.S. Vanyashin and M.V. Terentyev, Zh. Eksp. Teor. 
Fiz.  {\bf 48}, 565 (1965). 
 
\bibitem{Popov} M.S. Marinov and V.S. Popov, Yad. Fiz. 
\textbf{15}, 1271 (1972). 
 
\bibitem{Most} V.M. Mostepanenko, F.M. Frolov, and V.A. Sheliuto, 
Yad. Fiz. \textbf{37}, 1261 (1983). 
 
\bibitem{Kruglov} S.I. Kruglov, Int. J. Theor. Phys. \textbf{40}, 515 (2001). 

\bibitem{DB} D. Blaschke, V.N. Pervushin, D.V. Proskurin,  S.I. Vinitsky, and A.A. Gusev, Phys. of Atomic Nucl. 
\textbf{67}, 1050 (2004). 
 
 
\bibitem{Grib82} A.A. Grib and A.V. Nesteruk, Yad. Fiz. 
\textbf{35}, 216 (1982). 
  
\bibitem{Skalozub} V.V. Skalozub, Yad. Fiz. \textbf{21}, 
1337 (1975); \textbf{31}, 1980 (1980). 

\bibitem{rapp} E.~L.~Bratkovskaya, W.~Cassing, R.~Rapp and J.~Wambach,
Nucl.\ Phys.\ A {\bf 634}, 168 (1998), and references therein.
 
\bibitem{Smol} S.M. Schmidt, D.~Blaschke, G.~R\"opke, S.A.~Smolyansky, 
A.V.~Prozorkevich, and V.D.~Toneev,  Int.~J.~Mod.~Phys. E 
\textbf{7}, 709 (1998). 
 
\bibitem{OR} V.N. Pervushin, V.V. Skokov, A.V. Reichel, 
S.A. Smolyansky, and A.~V.~Prozorkevich, Int J. Mod. Phys. A (in 
press); hep-ph/0307200. 
 
\bibitem{Perv02} V.N. Pervushin, D.V. Proskurin, and A.A. Gusev, 
Grav. Cosmology \textbf{8}, 181 (2002). 
 
\bibitem{Zeld61} Ya.B. Zeldovich, JETP \textbf{41}, 1609 (1961) 
 
\bibitem{Smol03} S.A. Smolyansky, A.V. Reichel, 
D.V.~Vinnik, and S.M.~Schmidt, in \textit{Progress in 
Nonequilibrium Green's Functions II},  M. Bonitz and D. 
Semkat (eds.), (World Scientific, Singapore, 2003), p. 384. 

\bibitem{BS} C. Itzykson and J.B. Zuber, \textit{Quantum Field 
Theory}, (McGraw-Hill, 1980); \\ 
N.N. Bogoliubov and D.V. Shirkov, {\it Introduction to the 
Theory of Quantized Fields}, 3rd ed. (Wiley, 1980). 
 
\bibitem{Schmidt99} S. Schmidt, D.~Blaschke, G.~R\"opke, 
A.V.~Prozorkevich, S.A.~Smolyansky, and V.D.~Toneev, Phys. 
Rev. D \textbf{59}, 094005 (1999). 
 
\bibitem{18} A. Dolgov, in \textit{Multiple Facets of Quantization and 
Supersymmetry}, M. Olshavetsky, and A. Vainshtein (eds.),
(World Scientific, Singapore, 2002), p. 104.
 
\bibitem{Vinnik01} D.V. Vinnik, V.A. Mizerny, V.A. Prozorkevich, 
S.A. Smolyansky, and V.D. Toneev, Yad. Fiz. \textbf{64}, 
836 (2001). 
 
\bibitem{BD} 
D. Blaschke and M.P. D\c{a}browski, hep-th/0407078. 
 
\bibitem{Komatsu} E. Komatsu and  D.N. Spergel, Phys.Rev. D63, 063002 (2001); 
F. Vernizzi, A. Melchiorri, and R. Durrer, Phys. Rev. D63, 
063501 (2001). 
 
\bibitem{SmolHP} S.A. Smolyansky, A.V. Prozorkevich, D.V. Vinnik, 
and A.V. Reichel, Proc. of Int. Workshop "Hot point in 
Astrophysics", Dubna, 2000, p.364. 
 
\bibitem{23}  S.G. Mamaev and N.N. Trunov, Yad. Phys. 
\textbf{30}, 1302 (1979) (in russian). 
 
\bibitem{nw} Ya.B. Zeldovich and A.A. Starobinsky, Sov. Phys. 
JETP, \textbf{34}, 1159 (1971). 
 
 
\bibitem{19} J. Rau and B. M\"{u}ller, Phys. Rep. 
\textbf{272}, 1 (1996). 
 
\bibitem{20} M.P. Dabrowski and J. Stelmach, Ap. J. \textbf{97}, 978 (1989). 
 
\bibitem{25} J.A. Casas, J.Garcia-Bellido and M. Quiros, Class. 
Quant. Grav., \textbf{9}, 1371 (1992); G.W. Anderson and 
S.M. Carroll, astro-ph/9711288. 
 
 
\bibitem{22} P.B. Green and L. Kofman, Phys.Lett. B, \textbf{448}), 6 (1999). 
 
 
\bibitem{24} Proceedings of the 18-th IAP Astrophys. 
Colloquium ``On the Nature of Dark Energy'', Jul 1-5 (2002), 
Eds: Ph. Brax, J. Martin, J.-Ph. Uzan (Frontier Group, 
Paris, 2002). 
 
\bibitem{Parker} L. Parker, Phys. Rev. Lett. \textbf{21}, 562 (1968); 
Phys Rev. \textbf{183}, 1057 (1969); ibid, \textbf{D3}, 346 
(1971); Phys. Rev. Lett. \textbf{28}, 705 (1972); Phys. 
Rev. \textbf{D7}, 976 (1973). 

\bibitem{GM} A.A. Grib and S.G. Mamaev, 
Yad. Fiz. {\bf 10}, 1276 (1969).

\bibitem{SU} R.U. Sexl and H.K. Urbantke, Phys. Rev. {\bf 179}, 1247 (1969). 

\bibitem{Z} Ya.B. Zeldovich, Pisma Zh. Exp. Teor. Fiz. {\bf 12}, 443 (1970);  
 Ya.B. Zeldovich and A.A. Starobinsky,  Zh. Exp. Teor. Fiz. {\bf 61}, 2161
(1971).

\bibitem{Birrel} N.D. Birrell, P.C.W. Davies, \textit{Quantum Fields in Curved 
Spaces}, (Cambridge Univ. Press, Cambridge, 1982). 
 
\bibitem{ZN} Ya.B. Zeldovich and I.D. Novikov, \textit{Structure and 
Evolution of the Universe}, (Univ. Chicago Press, Chicago, 
1983). 
 
\bibitem{FM} Y. Fujii, K.-I. Maeda, \textit{The Scalar-Tensor 
Theory of Gravitation}, (Cambrige Univ. Press, Cambridge, 
2003). 
 
\end{thebibliography}
\end{document}